\RequirePackage{lineno}
\documentclass[aps,prb,preprint,groupedaddress,showkeys]{revtex4}
\usepackage[pdftex]{graphicx}
\usepackage[colorlinks=true,linkcolor=blue,citecolor=blue]{hyperref}
\graphicspath{{figures/}}
\DeclareGraphicsExtensions{.pdf,.jpeg,.png}

\usepackage{hyperref}
\usepackage{url}
\usepackage{multirow}
\usepackage{booktabs}
\usepackage{amsfonts}
\usepackage{amsmath}

\usepackage{enumitem}% custom enumerate
\usepackage{comment}

\begin{document}

\title{Mask Enhanced Deeply Supervised Prostate Cancer Detection on B-mode Micro-Ultrasound}

\author{Lichun Zhang$^{1}$, 
Steve Ran Zhou$^{2}$, 
Moon Hyung Choi$^{3}$, 
Jeong Hoon Lee$^1$, 
Shengtian Sang $^1$, 
Adam Kinnaird $^4$
Wayne G. Brisbane $^5$,  
Giovanni Lughezzani $^{6,7}$,
Davide Maffei $^{6,7}$,
Vittorio Fasulo $^{6,7}$, 
Patrick Albers $^4$,  
Sulaiman Vesal $^1$,
Wei Shao $^{8}$,
Ahmed N. El Kaffas $^{9}$,
Richard E.\ Fan$^2$, 
Geoffrey A. Sonn$^{1,2,\S}$, Mirabela Rusu$^{1,2,10,\S,*}$
}

\affiliation{$^{1}$Department of Radiology, Stanford University, Stanford, CA, USA} 
\affiliation{$^{2}$Department of Urology, Stanford University, Stanford, CA, USA}
\affiliation{$^{3}$The Catholic University of Korea, Department of Radiology, College of Medicine, 222 Banpo-daero Seocho-gu, Seoul, Republic of Korea, 06591}
\affiliation{$^{4}$Department of Urology, University of Alberta, Edmonton, Alberta, Canada }
\affiliation{$^{5}$Department of Urology, University of California Los Angeles, Los Angeles, California }
\affiliation{$^{6}$Department of Biomedical Sciences, Humanitas University, Pieve Emanuele, Milan, Italy
}
\affiliation{$^{7}$Department of Urology, IRCCS Humanitas Research Hospital, Milan, Italy}
\affiliation{$^{8}$Department of Medicine, University of Florida, Gainesville, Florida}
\affiliation{$^{9}$Department of Radiology, University of California San Diego, San Diego, California }
\affiliation{$^{10}$ Department of Biomedical Data Science }

\affiliation{$^{\S}$ Equal contribution as senior authors}
\affiliation{$^{*}$ Corresponding authors, email: mirabela.rusu@stanford.edu}

\begin{abstract}	% <= 500 words

\textbf{Background:} Prostate cancer is a leading cause of cancer-related deaths among men. Accurate detection of prostate cancer on radiology images is essential for early diagnosis, facilitating guidance for prostate biopsies and subsequent treatment planning. The recent development of high frequency, micro-ultrasound imaging offers improved resolution compared to conventional ultrasound and potentially a better ability to differentiate clinically significant cancer from normal tissue. However, the features of prostate cancer remain subtle, with ambiguous borders with normal tissue and large variations in appearance, making it challenging for both machine learning and humans to localize it on micro-ultrasound images.

\textbf{Purpose:} We sought to develop a comprehensive pipeline to process micro-ultrasound images and segment clinically significant prostate cancer on Brightness (B)-mode micro-ultrasound. To address the challenges associated with this task, we propose a novel \textbf{M}ask \textbf{E}nhance \textbf{D}eeply-supervised \textbf{M}icro-\textbf{US} network, termed MedMusNet, to automatically and more accurately segment prostate cancer to be used as potential targets for biopsy procedures. 

\textbf{Methods:} We used Magnetic Resonance Images (MRI) as reference to label micro-ultrasound exams, by mapping the pathology-confirmed radiology lesions onto the pseudo-sagittal plane of the micro-ultrasound. These labels were further refined by an expert clinician to match the micro-ultrasound image information. Our study included 64 men, 22 with normal or indolent cancer (ISUP Grade group 1) and 42 with clinically significant prostate cancer (Grade Group $\ge2$).

To ensure the topology consistency between 2D frames, we developed MedMusNet which takes as input only the B-mode micro-ultrasound images, detects and segments clinically significant cancer. MedMusNet leverages predicted masks of prostate cancer to enforce the learned features layer-wisely within the network, reducing the influence of noise and improving overall consistency across frames.

\textbf{Results:} MedMusNet successfully detected 76\% of clinically significant cancer with a Dice Similarity Coefficient of $0.365$, significantly outperforming the baseline Swin-M2F in specificity and accuracy (Wilcoxon test, Bonferroni correction, $p-value<0.05$). While the lesion-level and patient-level analyses showed improved performance compared to human experts and different baseline, the improvements did not reach statistical significance, likely on account of the small cohort. 

\textbf{Conclusion:} We have presented a novel approach to automatically detect and segment clinically significant prostate cancer on B-mode micro-ultrasound images. Our MedMusNet model outperformed other models, surpassing even human experts. These preliminary results suggest the potential for aiding urologists in prostate cancer diagnosis via biopsy and treatment decision-making. 
\end{abstract}

\maketitle

%\linenumbers\modulolinenumbers[5]

\section{Introduction}

% Prostate, Prostate Cancer, TRUS, biopsies
Prostate cancer is the second leading cause of cancer-related death among men in the United States  \cite{siegel2024cancer}. Early diagnosis significantly improves the 5-year survival rate for prostate cancer \cite{islami_annual_2021}. Currently, the standard diagnostic method for prostate cancer is histopathology analysis of samples obtained from ultrasound-guided prostate biopsies \cite{sarkar2016review}. However, due to the subtle features of prostate cancer on ultrasound images, ultrasound-guided biopsies are typically systematic, involving uniform sampling 12-14 regions of the prostate, without targeting suspicious regions as urologists are not confident in identifying lesions on conventional ultrasound \cite{harvey2012applications}.
Such blind systematic biopsies can miss 27-52\% cancers while carrying biopsy-related morbidities, e.g., pain, infection, sepsis \cite{loeb2013systematic, ahmed2017diagnostic}.
While Magnetic Resonance images (MRI) provide superior soft tissue contrast allowing the more reliable identification of prostate cancer\cite{ahmed2017diagnostic}, MRI often remains inaccessible due to its prohibitive cost and lack of access. 
Consequently, 64\% of patients undergo biopsy relying solely on conventional ultrasound\cite{Soerenson_Trends_2024}. 
Therefore, it is highly desirable to directly identify biopsy targets while balancing accuracy, efficiency, access, and cost-effectiveness.

\begin{figure}[htp]
\centering
\includegraphics[width=\linewidth]{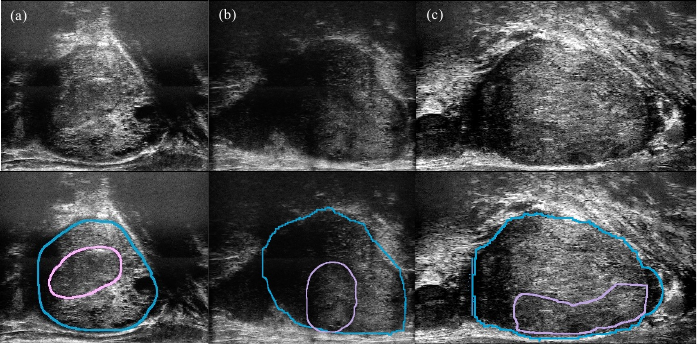}
\caption{ (\textbf{Top}) Examples of micro-ultrasound prostate images of three patients, (\textbf{Bottom}) depicting the prostate boundary (blue) and clinically significant cancer (purple). 
} \label{fig:micro-US}
\end{figure}

% 2. MicroUS, current segmentation methods \\
Recently, micro-ultrasound imaging has emerged as a promising imaging technology for the prostate \cite{rohrbach2018high}, offering improved resolution compared to conventional ultrasound (Fig.~\ref{fig:micro-US}). 
Micro-ultrasound was shown to have high sensitivity in detecting prostate cancer and in some studies similar accuracy as MRI \cite{sountoulides2021micro, dias2022multiparametric}. 
%Due to its real-time visualization capability and the elimination of the need for MRI-ultrasound fusion and its associate registration errors \cite{williams2022does}. Consequently, 
However, its interpretation remains challenging, with urologists having only 30\% agreement in localizing clinically significant lesions \cite{zhou_inter-reader_2024} with a sensitivity of 66\%. Such interpretation shortcomings may be remediated by using artificial intelligence methods, that have shown great promise in detecting prostate cancer on MRI \cite{saha_end--end_2021, seetharaman_automated_2021, bhattacharya_corrsignia_2021, saha_artificial_2024}. 

%============================================ Begin Figure
%============================================ End Figure

% Related work
Deep learning methods have revolutionized medical image segmentation, achieving impressive performance across various tasks \cite{lecun2015deep, litjens2017survey}. Numerous automatic prostate cancer segmentation and detection methods exist across different imaging modalities, e.g. MRI \cite{ozer2010supervised, duran2022prostattention}, MRI-pathology \cite{bhattacharya2022selective}, temporal-enhanced ultrasound \cite{moradi2008augmenting}, or conventional ultrasound \cite{rusu_procusnet_2024}. However, detecting prostate cancer on micro-ultrasound is challenging, and recent studies focused on using raw data (radio-frequency ultrasound signal), and weak labels offered by the biopsy without precise localization of cancer. Gilany \textit{et al.} \cite{gilany2022towards} proposed a deep model based on co-teaching \cite{han2018co} that improves the robustness of the approach to the label noise and employed an evidential deep learning method for uncertainty estimation. 
%They evaluated their method on a dataset of micro-ultrasound images from 194 patients who underwent prostate biopsy, achieving $67.38\pm4.91$ sensitivity, $88.20\pm6.85$ specificity and $87.76\pm1.82$ AUC (area under the curve). 
To address the issue of labeled data scarcity, Wilson \textit{et al.} \cite{wilson2023self} utilized a self-supervised learning method \cite{bardes2021vicreg} to extract features from large volumes of weakly labeled micro-ultrasound data. Their approach relied on transfer learning and detected cancer in the needle track using Radio-frequency ultrasound data from 1028 biopsy cores (n=391 subjects from two centers). Building on this approach,  Gilany \textit{et al.} \cite{gilany2023trusformer} utilized a similar method as a feature extractor and treated a set of regions of interest (ROIs) as a single biopsy core. They employed multiple instance learning \cite{dietterich1997solving, ilse2018attention} to aggregate ROI features and predict the tissue type for the entire biopsy core. 
%The model was trained and tested on a dataset comprising micro-ultrasound data from 578 patients who underwent prostate biopsy, achieving an AUROC of 80.3 \%. 
Recently, foundation models were also utilized to facilitate the classification of prostate cancer on the needle track  \cite{wilson_prostnfound_2024}.
However, these methods rely solely on weak labels provided by the biopsy to the rectangular region encompassing the biopsy needle and typically frame the cancer detection problem as a classification within selected small ROI that lie within the needle track. Consequently, while these methods can confirm the presence of cancer in a selected region, they fail to automatically and accurately localize cancer across the entire regions of micro-ultrasound images, potentially limiting their diagnostic utility, especially in providing targets for the biopsy.
%A recent study has sought to use nnUnet to segment cancer and the prostate on micro-ultrasound
%FIXME: How about steve's paper 

Recently, a few studies have shown the utility of B-mode micro-ultrasound, either in segmenting the prostate \cite{jiang_microsegnet_2024} or in creating ground truth labels of prostate cancer in patients undergoing radical prostatectomy by registering histopathology images into pseudo-whole mounts and subsequently onto micro-ultrasound \cite{imran_image_2023}. While the later approach can provide accurate cancer outlines for training AI models, it remains restricted to patients who undergo surgery, which all have clinically significant cancer confirmed by biopsy prior to the surgery.

% Challenges
Prostate cancer detection on micro-ultrasound images remains challenging for three reasons. First, micro-ultrasound captures series of 2D images in the pseudo-sagittal plane, whereas conventional processing, including expert annotation, is typically performed in the axial plane on MRI or whole mount histopathology images. This can make expert annotation of cancer in micro-ultrasound images laborious and error-prone. 
Second, micro-ultrasound images have speckle noise and artifacts caused by calcification and anatomic boundaries. Third, the appearance of cancer in micro-ultrasound images is highly heterogeneous often having indistinguishable borders with normal tissues (Fig.~\ref{fig:micro-US}).

% Summary
In this study, we developed a comprehensive pipeline to process micro-ultrasound images acquired in 2D pseudo-sagittal planes, and label clinically significant prostate cancer to allow the training of Artificial Intelligence (AI) methods on Brightness (B)-mode ultrasound. We introduced a novel \textbf{M}ask \textbf{E}nforced \textbf{D}eeply-supervised \textbf{M}icro-\textbf{U}ltra\textbf{U}trasound network, termed MedMusNet, to automatically and more accurately segment clinically significant prostate cancers as potential targets for biopsy procedures. 
%To the best of our knowledge, this is the first automated approach to segment PCas across the entire regions of micro-ultrasound images. 
The contributions of our work can be summarized as follows:

\begin{enumerate}
    \item We developed a comprehensive pipeline to process acquired micro-ultrasound images, facilitating expert annotations of prostate cancer both in the cartesian and cylindrical micro-ultrasound space.
    \item We proposed a novel mask-enhanced deep supervision mechanism that enhances the learned features layer-wisely within the network, thereby alleviating the adverse effects of image noise and improving the quality of the final segmentation.
    \item We extensively evaluated our method on a micro-ultrasound dataset and demonstrated superior performance compared to existing models and human readers both at Dice Similarity score and clinically focused pixel-level and lesion-level evaluations. 
\end{enumerate}

\section{Methods}

\subsection{Population Characteristics}
\label{s.dataset}

\textbf{Cohort:} We conducted an Institutional Review Board (IRB)-approved prospective collection of micro-ultrasound images from patients undergoing both diagnostic micro-ultrasound and subsequent conventional ultrasound-MRI fusion guided biopsy. 
The study included $64$ patients who underwent both MRI and micro-ultrasound procedures at Stanford University from 2022 to 2023. All patients underwent trans-perineal targeted biopsy of lesions with PI-RADS v2.1 scores $\ge$ 3. A 12-14-core systematic biopsy was performed to complement the targeted biopsy.

\textbf{Micro-ultrasound acquisition:}
Our dataset comprises $10,236$ 2D micro-ultrasound images.
During the micro-ultrasound scanning process, sequential 100 to 200 2D images were captured in a trans-rectal rotational manner. The 2D image size was $1372 \times 833$ and the image spacing was $0.03 mm \times 0.03 mm$.

\textbf{MRI acquisition:} The subjects included in this study, underwent 3 Tesla Magnetic Resonance Imaging (MRI) of the prostate prior to the biopsy. The exam included multiple sequences, T2 weighted MRI, Diffusion Weighted imaging, and the derived Appearance Diffusion Coefficient, Dynamic Contrast Enhance MRI and were interpreted as part of clinical care by board-certified radiologists. 

\textbf{MRI Labels:} Radiologists outline cancer on MRI for clinical care (PIRADS Scores $\ge 3$), to guide the conventional ultrasound-MRI fusion guided biopsy. The ground truth was cross-referenced by a urologist (SZ) and genitourinary radiologist (MH) jointly. They annotated areas on micro-ultrasound corresponding to biopsy-proven ISUP Grade Group (GG) $\geq$ 2 prostate cancer using 3D Slicer \cite{fedorov20123d}. 

\textbf{Multi-reader Labels:} In a previous study \cite{zhou_inter-reader_2024} we investigated the inter-reader variability  of six urologists (four institutions) interpreting micro-ultrasound (a subset of our cohort). All had completed a comprehensive online micro-ultrasound training program on the PRI-MUS protocol \cite{ghai2016assessing}. They were provided with image stacks for all patients and asked to annotate any suspicious lesions (PRI-MUS score $\geq$ 3) using 3D Slicer.

\begin{table}[]
    \centering
    \begin{tabular}{|l|cc|}
        \hline
        Variable  & Median/Value & IQR/percentage \\ \hline
        Number of Subjects & 64 & \\ 
        Number of Clin. Sign. Cancer & {\color{red}51} & \\
        Age & 69 & (61-73) \\ 
        PSa (ng/mL) & 6.1 & 4.9-8.7 \\ 
        Prostate Volume (cm$^3$) & 42.0 & 33.0-52.3 \\
        Number of Frames & 151 & 127-192 \\ \hline
        
        Patient Grade Group & &  \\
        ~~~ 0 & 9  & 14.1\% \\
        ~~~ 1 & 13 & 20.3\% \\
        ~~~ 2 & 18 & 28.1\% \\
        ~~~ 3 & 15 & 23.4\% \\
        ~~~ 4 & 2  &  3.1\% \\
        ~~~ 5 & 7  & 10.9\% \\ \hline
        Reference Lesion Count & & \\
        ~~~ 0 & 22  & 34.4\% \\
        ~~~ 1 & 34  & 53.1\% \\
        ~~~ 2 &  7  & 10.9\% \\
        ~~~ $>$2&1  & 1.6\% \\ \hline
        Reference lesion location & & \\
        ~~~ Anterior & 20 & 31.3 \\
        ~~~ Posterior& 44 & 68.8 \\ \hline
    \end{tabular}
    \caption{Cohort description. Abbreviation: IQR - inter-quartile range, Clin. sign - Clinically significant cancer.}
    \label{tab:my_label}
\end{table}

\subsection{Pre-processing}

%============================================ Begin Figure
\begin{figure}[hbp]
\centering
\includegraphics[width=0.8\linewidth]{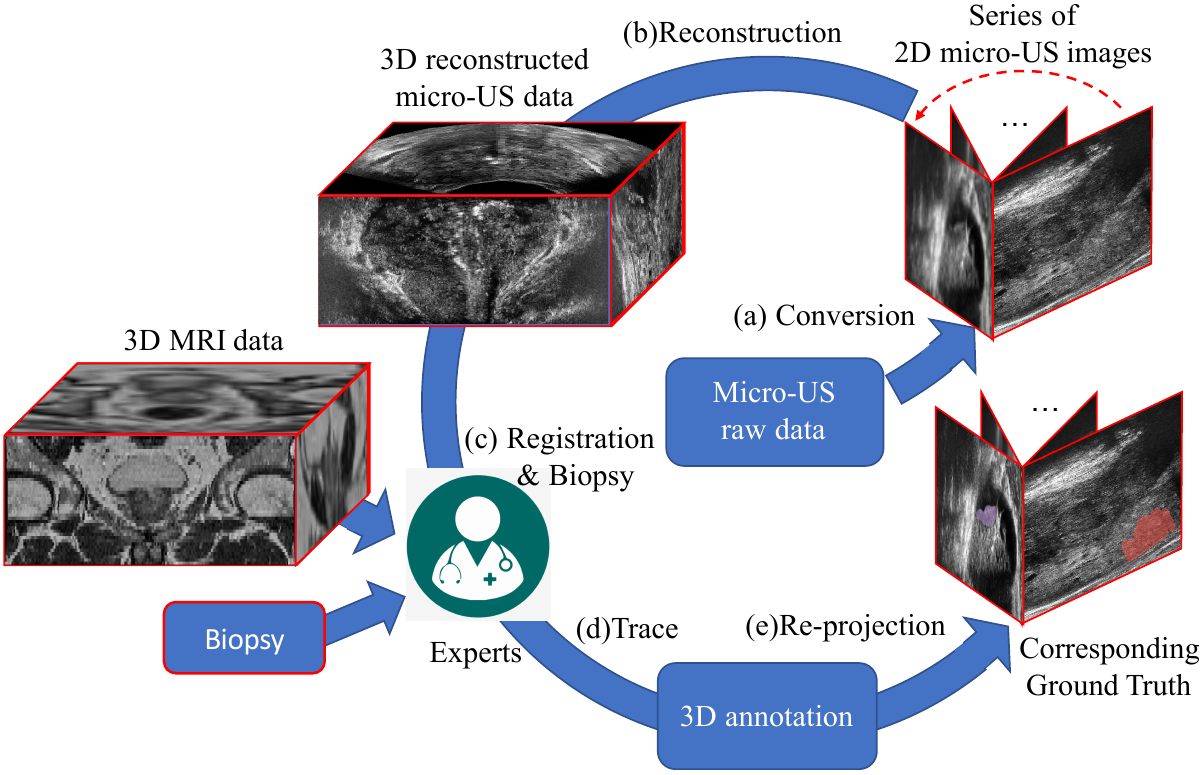}
\caption{Pre-processing pipeline to facilitate the labeling of prostate cancer from MRI onto micro-ultrasound. The steps involve the reconstruction of a Cartesian 3D micro-ultrasound volume, to which the MRI is registered allowing the projection of labels from MRI onto micro-ultrasound in the native pseudo sagittal space. 
} \label{fig:pipeline}
\end{figure}
%============================================ End Figure

The primary goal of data preprocessing is to accurately label prostate cancer in the native (cylindrical - pseudo-sagittal) coordinates of the micro-ultrasound images using biopsy-confirmed lesions originally outlined on MRI by radiologists. We developed a preprocessing pipeline involving several steps (\figurename~\ref{fig:pipeline}):
\begin{enumerate}[label=\textbf{Step (\alph*)}: , itemindent=1.2cm]
    \item We acquired a stack of 2D micro-ultrasound images covering the prostate from left to right (see Section \ref{s.acq}, \figurename~\ref{fig:scan}). 
    \item We reconstructed the series of 2D pseudo-sagittal micro-ultrasound images \cite{imran_image_2023} into the 3D Cartesian coordinates to facilitate spatial co-registration with MRI, a reference space where radiologists provide lesions targeted by the biopsy. 
    \item Manual Registration of the reconstructed micro-ultrasound images and MRI using affine transformations in 3D Slicer \cite{fedorov_3d_2012}  
    \item Expert clinician (SZ) delineates prostate cancer area using the MRI label reference (Section \ref{s.annotate}).
    \item Re-project the generated 3D cancer labels onto the 2D pseudo-sagittal micro-ultrasound images to serve as the ground truth for training AI models.

\end{enumerate}

\subsubsection{Image Acquisition}
\label{s.acq}
\begin{figure}[htp]
\centering
\includegraphics[width=0.8\linewidth]{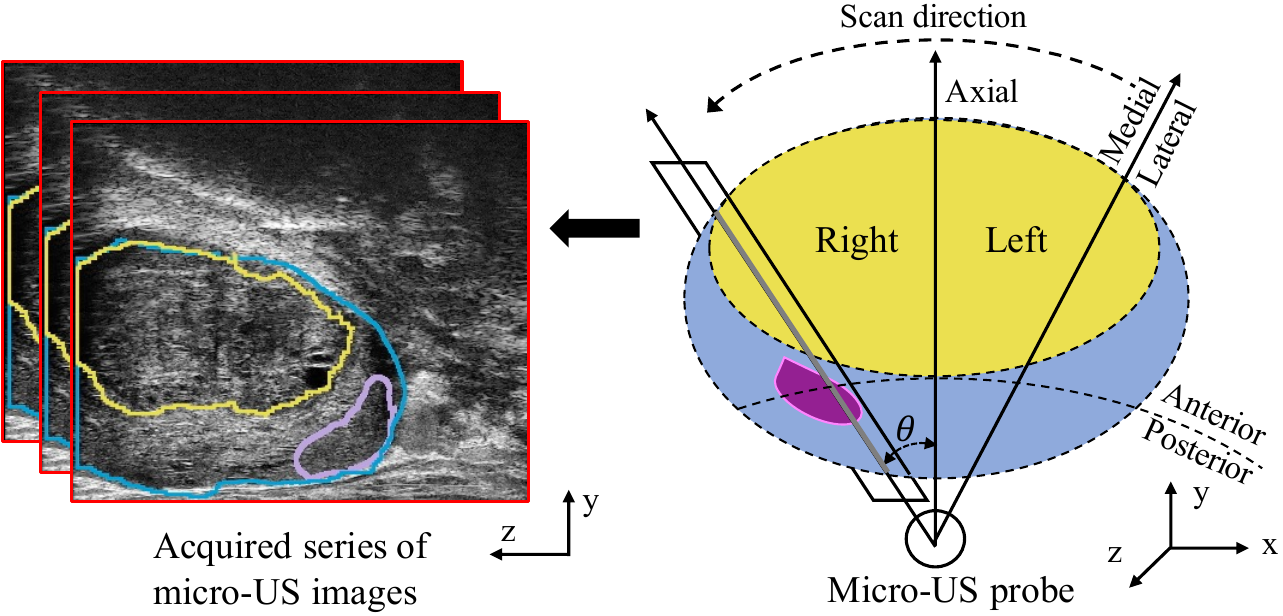}
\caption{Illustration of micro-ultrasound prostate scanning.  The transrectal micro-ultrasound probe rotates from left to right, generating a series of 2D images positioned in the pseudo-sagittal oblique plane, separated by the rotation angle $\theta$. Yellow indicates the transition zone, blue represents the peripheral zone, and purple denotes cancer lesions. 
} \label{fig:scan}
\end{figure}

The micro-ultrasound images of the prostate were obtained using the ExactVu transrectal micro-ultrasound system (Exact Imaging Inc., Markham, ON, Canada). During the prostate scanning process, the transrectal imaging probe was kept stationary in the cranial-to-caudal orientation and rotated from left to right, capturing a series of 100-300 micro-ultrasound 2D images (\figurename~\ref{fig:scan}). These images were positioned in the pseudo-sagittal oblique plane and separated by the rotation angle $\theta$. After the scan acquisition, the raw B-mode data were converted to a 0-255 range B-mode image and exported into a DICOM series using a customized MATLAB script. The rotation angle of each image was stored as the slice location in the DICOM file to facilitate subsequent reconstruction.

%============================================ Begin Figure
%============================================ End Figure

\subsubsection{Image Annotation}
\label{s.annotate}
%============================================ Begin Figure
\begin{figure}[htp]
\centering
\includegraphics[width=\linewidth]{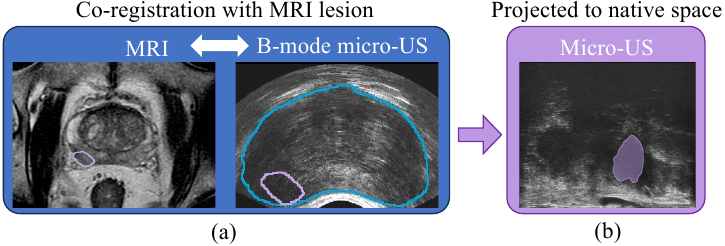}
\caption{Illustration of mapping the location of clinically significant cancer (ISUP Grade Group $\geq 2$) on micro-ultrasound images using MR images. (a) The MRI is manually registered using affine transformations to the 3D Cartesian micro-ultrasound scan using 3D Slicer guided by the prostate boundary (Blue). The radiologists' annotations of biopsy-confirmed MRI-visible lesions (purple) are projected from MRI onto 3D Cartesian B-mode micro-ultrasound images. (b) The ground truth cancer labels (purple) are projected back onto the native pseudo-sagittal images, and refined by export in the native space to reduce the effect of registration errors and interpolation artifacts.
} \label{fig:mapping}
\end{figure}
%============================================ End Figure

%repetitiove and doesn't add more infomrmation
The creation of ground truth labels for clinically significant cancer can be challenging due to the subtle differences between cancer and normal tissues, and the shadowing artifacts caused by calcification and anatomic boundaries. 
%While the biopsy track can provide weak labels, they fail to provide the accurate location of cancer across the entire region. 
To address these challenges, our pre-processing pipeline used MRI, and corresponding biopsy-confirmed radiologist lesions. This facilitates the identification of suspicious lesions and the generation of strong labels in micro-ultrasound images. 
To establish pixel correspondences between the MRI and micro-ultrasound, we developed a reconstruction algorithm that enables forward and backward projections between the cylindrical sagittal plane of the native micro-ultrasound images and the Cartesian coordinates of MRI. Consequently, the MRI can be reliably registered to the reconstructed B-mode micro-ultrasound volume\ (\figurename~\ref{fig:pipeline} Steps b-c). The micro-ultrasound images, biopsy pathology and MRI results were cross-referenced by a urologist and a genitourinary radiologist to annotate areas on the B-mode micro-ultrasound volumes corresponding to biopsy-proven grade group (GG) $\geq 2$ prostate cancer (\figurename~\ref{fig:pipeline} Step d). Finally, the cancer labels were projected onto the native pseudo-sagittal micro-ultrasound images using the backward reconstruction  (\figurename~\ref{fig:pipeline} Step e, \figurename~ \ref{fig:mapping}).

\subsection{MedMusNet Architecture}

\begin{figure}[t!]
\centering
\includegraphics[width=\linewidth]{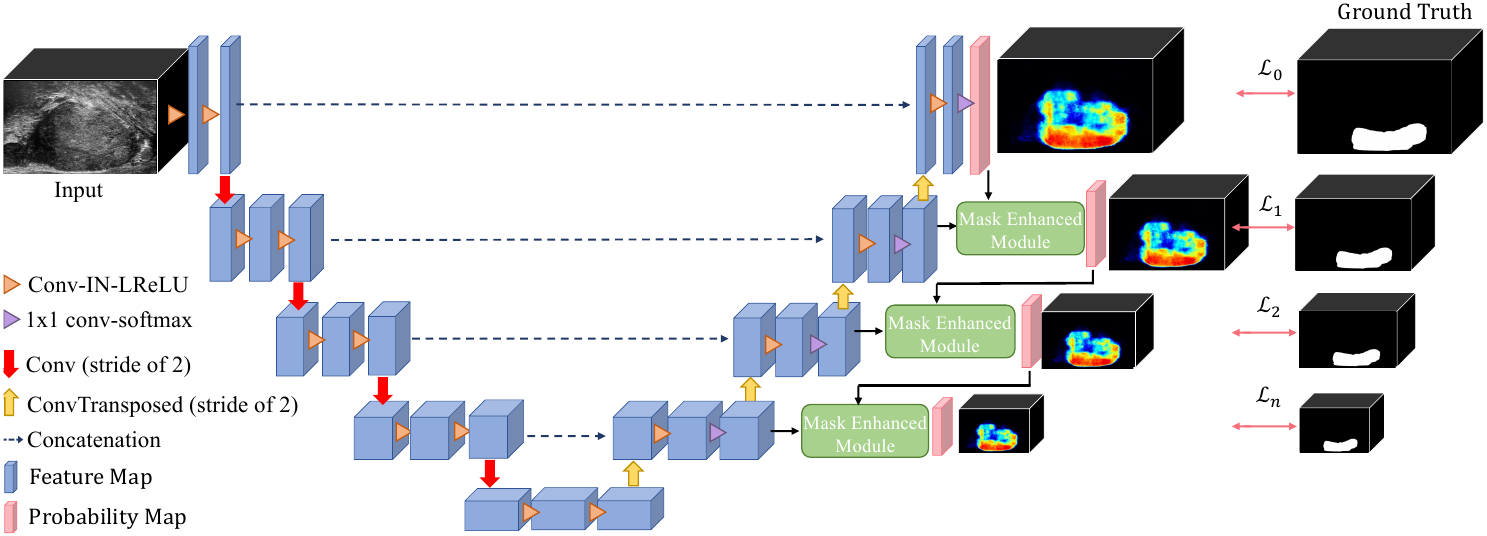}
\caption{Overview of the proposed MedMusNet model. Based on a 3D UNet-like backbone, the model incorporates the spatial relationship explicitly and enhanced global features captured by mask-enhanced modules and the multi-scale deep supervision. $\mathcal{L}_n$ represents multi-scale losses and the overall loss $\mathcal{L}$ is their weighted sum (Eq.\ \ref{eq:Ln} and Eq.\ \ref{eq:L}). 
} \label{fig:network}
\end{figure}

% 3D UNet
MedMusNet consists of a classical 3D UNet-like backbone \cite{cciccek20163d}, a mask-enhanced module, and a multi-scale deep supervision module (\figurename~\ref{fig:network}). Similar to the standard UNet, the backbone includes an encoder (left side) and a decoder (right side) each with multiple resolution layers. Each encoder layer comprises two convolutions followed by an instance norm (IN) unit and a leaky rectified linear unit (LReLU). Instead of the conventional fixed pooling operator, a convolution operator with a stride of two is used at each downsampling step to facilitate more effective feature learning. Each decoder layer contains a 3D transposed convolution operator (ConvTransposed) with a stride of two for upsampling, followed by a concatenation and two convolutions, each succeeded by an IN unit and a LReLU. Shortcut connections provide cropped high-resolution feature maps from the encoder to the corresponding up-sampled features from layers of equal resolution in the decoder. In the last layer, a $1 \times 1\times 1$ convolution unit reduces the number of output channels to match the number of labels and a softmax unit outputs the final probability map.  

\label{s.network}
%============================================ Begin Figure
%============================================ End Figure

\label{s.meds}
%============================================ Begin Figure
\begin{figure}[htb]
\centering
\includegraphics[width=0.8\linewidth]{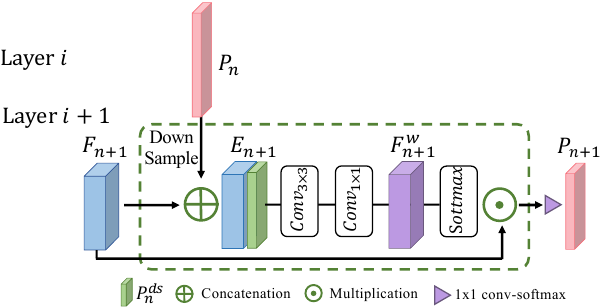}
\caption{Illustration of the mask enhanced module.
} \label{fig:meds}
\end{figure}
%============================================ End Figure

% Mask enhanced module 
However, the direct segmentation output from the backbone is often insufficiently precise and exhibits erroneous localization of lesions (e.g., over-segmentation outside the prostate), particularly in micro-ultrasound images that routinely incorporate speckle and artifacts. To mitigate the influence of noise and improve the segmentation accuracy, we proposed the \textbf{M}ask \textbf{E}nhanced \textbf{M}odule (MEM) and incorporated the multi-scale deep supervision. The mask-enhanced module concatenates the probability map from an intermediate layer with the high-resolution probability map from an upper layer in the decoder, aiming to incorporate the spatial constraints from a larger receptive field for better supervision (\figurename~\ref{fig:meds}). Moreover, the lower-scale features encompassing global information are effective integrated into the higher-scale features which include local contextual details, leading to a reduction in the impact of background noise and an enhanced emphasis on foreground features. Specifically, for the probability map, $P_n \in \mathbb {R}^{B\times C \times D_n \times H_n \times W_n} $ at the layer $n$, it was first down-sampled into $P_n^{ds} \in \mathbb {R}^{B\times 1 \times D_{n+1} \times H_{n+1} \times W_{n+1}} $ to match the size of the feature map $F_{n+1} \in \mathbb {R}^{B \times C_{n+1} \times D_{n+1} \times H_{n+1} \times W_{n+1}}$ at the layer $n+1$, where $B$ is the batch size, $C$ is the number of labels, and $D$, $H$ and $W$ represent the depth, height and width of a 3D image tensor, respectively. We obtained the feature embedding $E_{n+1} \in \mathbb {R}^{B\times (C_{n+1}+1) \times D_{n+1} \times H_{n+1} \times W_{n+1}} $ after the concatenation of $P_n^{ds}$ and $F_{n+1}$ along the channel dimension. A $Conv_{3\times3}$ and $Conv_{1\times1}$ are then utilized to process $E_{n+1}$ to generate attention weight feature $F_{n+1}^w \in \mathbb {R}^{B\times (C_{n+1}) \times D_{n+1} \times H_{n+1} \times W_{n+1} }$, and finally output the probability map $P_{n+1}$ at layer $n+1$:
% \begin{equation}
%     P_{n+1} = softmax(Conv_{1\times1}(F_{n+1} \odot softmax(F_{n+1}^w))) \;\;,
%     \label{eq:prob}
% \end{equation}

\begin{align}
    E_{n+1}   &= F_{n+1} \oplus \operatorname{DownSample}(P_n, 2^{-1})    \label{eq:e_n+1} \;\;, \\
    F^w_{n+1} &= Conv_{1\times1}(Conv_{3\times3}(E_{n+1}))                \label{eq:f_w}   \;\;, \\
    P_{n+1}   &= softmax(Conv_{1\times1}(F_{n+1} \odot softmax(F^w_{n+1})))   \label{eq:p_n+1}\;\;,
\end{align}
where $\odot$ denotes element-wise product.

% Deep supervision + loss function
To enhance the ability of the learned features to discriminate normal vs cancer and boost the model's ability to accurately segment cancer, we utilized deep supervision \cite{lee2015deeply} at multiple scales of the model. We computed the segmentation loss $\mathcal{L}_n, n \in [0, N-1]$ between the deep supervision output $P_n^{ds}$ and the corresponding ground truth $GT_n^{ds}$ at each layer $n$ in the decoder (\figurename~\ref{fig:network}). Due to the different resolutions of layers in the decoder, the ground truth for each layer is correspondingly down-sampled, and the weights are halved with each decrease in resolution. The loss function combines cross-entropy loss and Dice loss \cite{drozdzal2016importance}, with the objective to minimize the overall loss $\mathcal{L}$, the normalized weighted sum of $\mathcal{L}_n$ across all resolutions:
\begin{align}
    GT_n^{ds}     &= \operatorname{DownSample}(GT, scale=2^{-n}) \;\;,  \label{eq:gt_ds} \\
    \mathcal{L}_n &= \operatorname{DiceLoss}(P_n^{ds}, GT_n^{ds}) + \operatorname{CE}(P_n^{ds}, GT_n^{ds})  \label{eq:Ln} \;\;, \\ 
    \mathcal{L}   &= \frac{ \sum^{N-1}_{n=0}{ 2^{-n} \mathcal{L}_n}} {\sum^{N-1}_{n=0}{ 2^{-n}}}    \label{eq:L} \;\;,  
\end{align}
where $N$ is the number of layers of the network model, $Out_{ds_n}$ and $GT_{ds_{n}}$ represent the output and corresponding ground truth at the deep supervision layer $n$ ($n \in [0,N]$), respectively. This strategy allows the multi-scale deep supervision layers to act as proxies for assessing the quality of features at various levels. They guide the model to prioritize intermediate layers that produce highly discriminative feature maps, ultimately leading to improved performance of the model trained on these features.
 
\subsection{Post-processing}
% Post processing
We utilized a series of conventional post-processing methods to remove small false positives, primarily focusing on morphology and connected-component analysis. Initially, a closing operation with a kernel size of three pixels was applied to fill small holes and connect-closely disjoint predicted lesion areas by filling narrow gaps. During connect-component analysis, a threshold of $10,000$ pixels was set, ensuring that minor false positive areas smaller than this threshold were disregarded from the prediction.

\subsection{Evaluation metrics}

% Metrics 
We used the Dice similarity coefficient (DSC) as a measure of the overlap between the segmentation of MedMusNet and the ground truth labels. In response to experts' clinical needs to perform a more comprehensive evaluation of the performance of MedMusNet, we conducted a lesion and patient-level evaluation that includes the sensitivity, specificity, accuracy, positive predictive values (PPV), negative predictive values (NPV) and F1 score. Outlines from MedMusNet were deemed true positive if there was more than $20\%$ overlap with a ground truth lesion. Unlike lesion-level evaluation which focuses on identifying the individual lesion including multiple foci, patient-level evaluation emphasizes the ability to accurately identify any clinically significant prostate cancer lesion when present. A $100\%$ patient-level accuracy indicates that all ground truth cancer lesions were correctly identified.

% Sector evaluation is only used for reader agreement
% To comprehensively evaluate the segmentation performance, we conducted a sector-based analysis at both the lesion and patient level using statistical metrics including sensitivity, specificity, accuracy, positive predictive values (PPV), negative predictive values (NPV) and F1 score. As illustrated in \figurename~\ref{fig:eval-sec}, each prostate was computationally divided into $30$ sectors based on transperineal biopsy templates. A sector was considered to contain lesions if more than $20\%$ of an annotated volume fell within that sector, or-to account for large lesion-if more than $50\%$ of the sector volume was occupied by an annotation.

%============================================ Begin Figure
% \begin{figure}[htp]
% \centering
% \includegraphics[width=0.6\linewidth]{sector.pdf}
% \caption{TBD: sector-based evaluation \xxx{Need a 3D volume figure on the far right, just like Fig.2 B in Steve's European Urology paper.}
% } \label{fig:eval-sec}
% \end{figure}
%============================================ End Figure

\subsection{Statistical Analysis}

We performed non-parametric Paired Wilcoxon tests with Bonferroni multiple-comparison correction \cite{bonferroni_teoria_1936} to compare the sensitivity, specificity, NPV, PPV and Dice coefficients testing differences between our model and the baseline nnUNet or the average for the readers. Moreover, we used the non-parametric unpaired Mann-Whitney U test \cite{mcknight_mann-whitney_2010} with Bonferroni multiple-comparison correction to test the differences between the performance of MedMusNet for the anterior and posterior lesions.

\subsection{Implementation details and Experimental design}
% Implementation details
Five-fold cross-validation was used for train and evaluate the performance of the proposed method. Due to the initial resolution discrepancy between axes at the micro-ultrasound images, whose pseudo 'z-size' and 'z-spacing' is $100\sim200$ and $0.3 mm$ in 3D space, respectively, an image patch size of $32 \times 192 \times 256$ was set during the training of 3D models (i.e., nnUNet 3D, MusNet and MedMusNet) to ensure isotropic feature representation. The training settings included a batch size of 2, the number of layers $N=6$, a learning rate of $0.01$ with a momentum of $0.99$, and a weight decay of $3e^{-5}$. Kaiming initialization \cite{he2015delving} and Adam optimizer \cite{kingma2014adam} were employed, with the training duration set to $1,000$ epochs. These hyper-parameters were kept consistent across the 3D CNN-based models. For Swin-M2F, we followed the settings of the standard base-sized (Swin-B) model as outlined in \cite{liu2021swin, cheng2022masked}. Training and evaluation processes were conducted on an NVIDIA A6000 GPU with 48 GB memory. %All experiments were performed using Python 3.11.4 and PyTorch \cite{paszke2017automatic} version 2.0.1.

We compared the following models/readers across the n=37 subjects that overlap between our study and the one preported in \cite{zhou_inter-reader_2024}
\begin{itemize}
    \item Swin-M2F: The model employs the transformer-based Swin-Transformer \cite{liu2021swin} backbone along with Mask2Former \cite{cheng2022masked} decoders,
    \item nnUNet-3D: This represents the well established framework based on the uNet architecture, that optimized the hyperparameters based on the data signatures \cite{isensee2021nnu}, 
    \item MusNet: This models uses the MedMusNet Architectures without the mask enhanced module (MEM), which allows us to test the utility of the mask enhance module.
    \item Readers include the six clinicians that were part of the inter-reader variability study \cite{zhou_inter-reader_2024}, and reported results represent average performance metrics \cite{zhou2024inter}.
\end{itemize}

\section{Results}
% \subsection{Qualitative results}

%============================================ Begin Figure
\begin{figure}[htp]
\centering
\includegraphics[width=\linewidth]{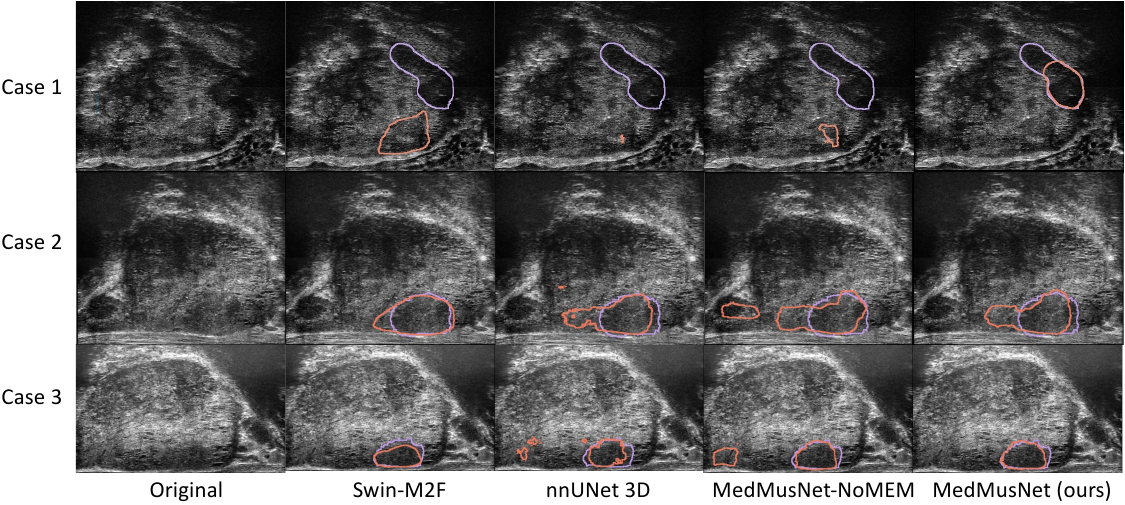}
\caption{Qualitative evaluation of our proposed approach overlaid on prostate micro-ultrasound images of three representative cases. \textbf{(Case 1)} large-sized clinically significance cancer spanning the peripheral and transition zones with indistinct lesion patterns. \textbf{(Case 2)} medium-sized lesion located in the peripheral zone with relatively clear boundaries. \textbf{(Case 3)} small-sized cancer in the peripheral zone with relatively clear boundaries. Each row displays a representative 2D frame from different patients and each column represents a compared method. Purple indicates the ground truth clinically significant cancer and the orange represents results from MedMusNet that successfully detected and segmented all three lesions.
} \label{fig:results}
\end{figure}
%============================================ End Figure

% %============================================ Begin Table
\begin{table*}[tp]
\centering
\caption{Lesion-level evaluation (n=37 subjects). The dice coefficient was not computed for the readers (last row) as they were not instructed to segment the entire lesion, but only representative slices due to the tedious nature of annotating cases. Hence the dice coefficient is not directly comparable with the models' outlines.}
\begin{tabular}{lcccccccc}
\toprule
               & No. Cases & \textbf{Sensitivity}   & \textbf{Specificity}   & \textbf{Accuracy}  & \textbf{PPV}  & \textbf{NPV} & \textbf{DSC}  \\
\midrule
%\textbf{MusNet}             &  37                & 0.74 & 0.95 & 0.94 & 0.31 & 0.99 & 0.318\\
%\textbf{nnUNet-3D}          &  37                 & 0.70 & 0.94 & 0.93 & 0.28 & 0.99 & 0.287\\
%\textbf{Swin-M2F}           & 37                  & 0.66 & 0.92 & 0.91 & 0.25 & 0.99 & 0.336\\
\textbf{Swin-M2F}           & 64                  & 0.61 & 0.88 & 0.87 & 0.24 & 0.99 & 0.283\\
\textbf{nnUNet-3D}          & 64                  & 0.69 & 0.92 & 0.92 & 0.25 & 0.99 & 0.236\\
\textbf{MusNet}             & 64                & 0.72 & 0.94 & 0.93 & 0.29 & 0.99 & 0.262\\
\textbf{MedMusNet (Ours)}   & 64                & 0.77 & 0.94 & 0.93 & 0.35 & 0.99 & 0.313\\
\textbf{MedMusNet (Ours)}   & 37                & 0.76 & 0.96 & 0.95 & 0.42 & 0.99 & 0.365\\
\textbf{Readers} \cite{zhou_inter-reader_2024} & 37 &0.58 & 0.98 & 0.97 & 0.51 & 0.99 & -  \\

\bottomrule
\end{tabular}
\label{table-lesion}
\end{table*}
% %============================================ End Table

%============================================ Begin Figure
\begin{figure}[htp]
\centering
\includegraphics[width=0.5\linewidth]{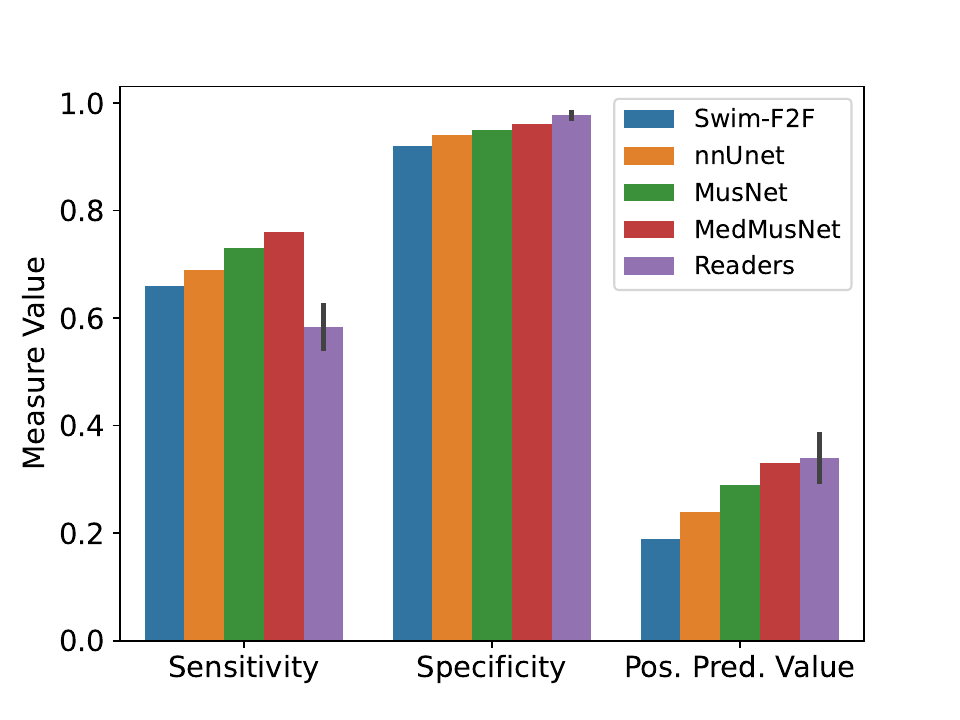}
\caption{Quantitative comparison between MedMusNet, other models and Readers (n=37 subjects).} \label{fig:results2}
\end{figure}
%============================================ End Figure
% Figure Explaination
\figurename~\ref{fig:results} presents the qualitative segmentation results of four different models across three representative cases. For Case 1, the micro-ultrasound scan displayed indistinct lesion patterns and considerable artifacts. Consequently, all models except ours failed to accurately identify the correct region, despite the presence of a large-sized prostate cancer spanning the peripheral and transitional zone. Benefiting from the incorporated mask enhanced module, our method, in contrast, successfully detected the approximate location of the lesion and segments out the major region of the lesion. For Cases 2 and 3, all models demonstrated relative accurate prostate cancer segmentation meanwhile while also displaying instances of over-segmentation and under-segmentation. Although our method exhibited these issues as well, it overall outperformed the other models. Notably, our method corrected major errors in the MusNet model after incorporating MEM, showcasing its accuracy and robustness against artifacts. The comparison between these two models further demonstrated the effectiveness of spatial constraints and multi-scale feature integration provided by the mask-enhanced module, enabling the model to effectively capture image features and focus on challenging regions.

% \subsection{Quantitative results}

% Tables and explains
%Table \ref{table-DSC} displays the five-fold cross-validation DSC scores for four different models. 
Our MedMusNet model exhibited superior performance in pixel-level clinically significant cancer segmentation compared to the Swin-M2F model and the nnUNet-3D model (Table~\ref{table-lesion}, Supplementary Table~\ref{table-patient}, \figurename~\ref{fig:results2}). DSC improvement from $0.262$ in MusNet to $0.313$ in MedMusNet validates the effectiveness of our proposed mask-enhanced module. 
%Tables \ref{table-lesion} and \ref{table-patient} present the statistic results of lesion-level and patient-level evaluation from five experiments, respectively. 
The MusNet model achieved overall better performance compared to other models, demonstrating the effectiveness of our network architecture design (Table~\ref{table-lesion}, Supplementary Table~\ref{table-patient}). The mask-enahnced module helped MedMusNet increase the lesion level PPV (Table~\ref{table-lesion}) and NPV in patient-level evaluation (Supplementary Table~\ref{table-patient}). It ultimately outperformed the other three models across most metrics. This improvement is also reflected in \figurename~\ref{fig:results}, where our MedMusNet effectively increased the true positive lesions (Case 1) and suppressed the false positive predictions (Cases 2 and 3). Notably, our method demonstrated overall performance close to that of the multiple expert readers and  finding more cancers as indicated by the increased sensitivity. This underscores the potential of our method to assist clinicians in accurately localizing prostate cancer during micro-ultrasound-guided biopsy. We compared the proposed model with the baseline Swin-M2F and found statistically significant improvements in specificity and accuracy (Wilcoxon test, Bonferroni correction, p$<0.05$), however statistical differences for the other metrics were not observed. 

\begin{comment}
% %============================================ Begin Table
\begin{table}[htp]
    \centering
    \caption{Comparison of DSC scores across different models}
    \begin{tabular}{ccccc}
    \toprule
    & \textbf{Swin-M2F} & \textbf{nnUNet-3D} & \textbf{MedMusNet-NoMEM} & \textbf{MedMusNet (ours)} \\
    \midrule
    \textbf{DSC} & 0.283 & 0.236 & 0.262 & 0.313 \\
    \bottomrule
    \end{tabular}
    \label{table-DSC}
\end{table}
% %============================================ End Table
\end{comment}

\subsection{Subgroup Analysis} 

We compared the performance of MedMusNet for Anterior vs Posterior lesions and, while we observe performance improvements as estimated by the lesion level Sensitivity (0.85 vs 0.64) and PPV (0.4 vs 0.26), these differences do not show statistically significant differences. The accuracy was statistically better for posterior lesions compared to anterior lesions (Table~\ref{table-lesion-anterior}), suggesting the value of the approach specifically for posterior lesions. Of note, the reader had worse ability to find anterior lesions, and while still models the MedMusNet might still be successful in assisting them in finding more of the anterior lesions. 

\begin{table*}[htp]
\centering
\caption{Lesion-level Evaluation for Anterior vs Posterior Lesions for predictions for our proposed model, MedMusNet (n = 64 subjects). Only the accuracy was statistically significantly better for Posterior tumors compared to anterior tumors (Mann-Whitney test, Bonferroni correction, p-value$<0.05$). All represent results for the 64 cases, while Anterior and Posterior rows represent aggregations of the results for the cases with lesions either in the anterior or posterior side of the prostate.}
\begin{tabular}{lcccccc}

               & \multicolumn{1}{c}{\textbf{Sensitivity}}   & \multicolumn{1}{c}{\textbf{Specificity}}   & \multicolumn{1}{c}{\textbf{Accuracy}}  & \multicolumn{1}{c}{\textbf{PPV}}  & \multicolumn{1}{c}{\textbf{NPV}}  \\
\midrule
% Miumbers for n = 37
%\textbf{All}       & 0.76 & 0.96 & 0.95 & 0.42 & 0.99  \\ 
%\textbf{Anterior}  & 0.43 & 0.93 & 0.92 & 0.19 & 0.98  \\
%\textbf{Posterior} & 0.89 & 0.96 & 0.96 & 0.50 & 1.00  \\ 
%Numbers for n = 64
\textbf{All}       & 0.77 & 0.94 & 0.93 & 0.35 & 0.99    \\ 
\textbf{Anterior}  & 0.64 & 0.92 & \textbf{0.91} & 0.26 & 0.99  \\
\textbf{Posterior} & 0.85 & 0.95 & \textbf{0.95} & 0.40 & 1.00   \\ 

\bottomrule
\end{tabular}
\label{table-lesion-anterior}
\end{table*}

\section{Discussion}
Our study introduced MedMusNet, a deep learning network that successfully detected and outlined the extent of clinically significant cancer on micro-ultrasound, an exciting new high-resolution ultrasound modality. MedMusNet ensures topology consistency in 3D at different resolutions to limit the effect of noise that are common in any ultrasound acquisitions. 
MedMusNet reliably identified the extent of clinically significant cancer, especially those located in the peripheral zone, improving the overlap with the lesion and sensitivity compared to other models and humans. 

While some studies have shown that expert urologists diagnosed the majority of cancers using micro-ultrasound \cite{lughezzani2021diagnostic}, others have found its performance lower than MRI and plagued by low inter-reader agreement (30\% \cite{zhou2024inter}). 
Compared to six experts, MedMusNet had an overall higher sensitivity, detecting 76\% lesions vs 58\% by readers (but did not reach statistical significance), at the cost of some false positives (0.96 vs 0.98 in lesion-level evaluation, 0.10 vs 0.39 in patient-level analysis). 
The difference in specificity between lesion- and patient-level evaluations is caused by the challenges in defining negative regions in a way that is comparable to a lesion outlined by clinicians (used for sensitivity calculation). Here we defined the negative regions based on the sectors used during the transperineal biopsy, while also dividing the prostate into the three subregions (apex, mid-gland, base), accounting for a total of 39 regions. While it can be straightforward to find sectors that are negative as ground truth and as MedMusNet predictions, overall at patient-level the MedMusNet has small false positives that make the predictions at patient-level less specific.    

Along with the development of MedMusNet, another significant contribution of the current study is the introduction of the comprehensive pipeline to process micro-ultrasound images and create ground truth for clinically significant prostate cancer on micro-ultrasound for patients with prior MRI. While our goal was to develop this approach as a strategy to label clinically significant cancer in the native pseudo sagittal space it may provide useful functionality for micro-ultrasound-MRI fusion-guided biopsy.

Two features distinguish this study from existing methods \cite{wilson2023self, gilany_trusformer_2023, wilson_prostnfound_2024}. First, we used B-mode micro-ultrasound images, without requiring radio-frequency data or other raw data which is typically unavailable outside research studies. Second, MedMusNet not only detects but also outlines the extent of clinically significant cancer on micro-ultrasound, unlike prior studies that rely on clinicians to identify regions that can be confirmed to have cancer, limiting its utility for targeting and artificially improving performance metrics.

Our study has three noteworthy limitations. 
First, the ultrasound images included in our study were collected from a single institution. Future studies will focus on including data from other institutions.
Second, the number of cases in this prospective study is limited. Prospective studies are particularly challenging often limiting the number of subjects to be include, and the power of our statistical analysis. While we continue recruiting subjects for our study, these preliminary results are encouraging, suggesting the utility of micro-ultrasound for targeting suspicious lesions.
Third, our model has a large number of false positives compared to human experts. While this comes with increased sensitivity, future studies will focus on reducing the number of false positives.

Despite these limitations, our approach holds promise in facilitating the identification of biopsy targets using micro-ultrasound, providing guidance in the absence of MRI, which remains unavailable in as many of 64\% of prostate biopsies in the USA \cite{Soerenson_Trends_2024} and considerably more globally. While micro-ultrasound is gaining popularity due to its higher resolution, automated lesion identification such as the one provided by MedMusNet will enhance its clinical utility. 

\section{Conclusion}
% TODO:
%
Our approach, MedMusNet successfully detected and segmented clinically significant prostate cancer on micro-ultrasound while taking advantage of mask-enhanced and deep supervision modules to reduce the effect of noise. MedMusNet outperformed alternative approaches, suggesting its clinical utility when assisting experts using micro-ultrasound to guide prostate biopsies or local treatment. Currently, targeting suspicious lesions for prostate cancer is only possible when MRI is available, however, micro-ultrasound combined with detection models like MedMusNet can help provide more readily available alternatives.   

\section{Acknowledgments} 

This work was supported by Stanford University (Departments: Radiology, Urology) and by National Cancer Institute, National Institutes of Health (R37CA260346). The content is solely the responsibility of the authors and does not necessarily represent the official views of the National Institutes of Health. 
 
% References

\newpage

\section{Supplementary Material}

\subsection{Supplementary Tables}

\renewcommand\thefigure{S\arabic{figure}}    
\setcounter{figure}{0}    
\renewcommand\thetable{S\arabic{table}}
\setcounter{table}{0} 

% %============================================ Begin Table
\begin{table*}[htp]
\centering
\caption{Patient-level evaluation}
\begin{tabular}{lcccccc}
\toprule
               & \multicolumn{1}{c}{\textbf{Sensitivity}}   & \multicolumn{1}{c}{\textbf{Specificity}}   & \multicolumn{1}{c}{\textbf{Accuracy}}  & \multicolumn{1}{c}{\textbf{PPV}}  & \multicolumn{1}{c}{\textbf{NPV}}   \\
\midrule
\textbf{MedMusNet (Ours)}& 0.77 & 0.10 & 0.56 & 0.66 & 0.14 \\
\textbf{MusNet}          & 0.76 & 0.00 & 0.53 & 0.63 & 0.00 \\
\textbf{nnUNet-3D}       & 0.72 & 0.10 & 0.53 & 0.64 & 0.13 \\
\textbf{Swin-M2F}        & 0.68 & 0.18 & 0.53 & 0.65 & 0.20 \\
\textbf{Readers}& 0.66 & 0.39 & 0.58 & 0.73 & 0.31 \\
\bottomrule
\end{tabular}
\label{table-patient}
\end{table*}
% %============================================ End Table

\end{document}